\documentclass[twocolumn]{jpsj2}

\usepackage{times}

\usepackage{graphicx}

\title{Multipole Susceptibility
of Multiorbital Anderson Model
Coupled with Jahn-Teller Phonons}

\author{Takashi {\sc Hotta}}

\inst{Advanced Science Research Center,
Japan Atomic Energy Agency,
Tokai, Ibaraki 319-1195}

\recdate{\today}

\abst{
In order to clarify possible multipole states of Sm-based filled
skutterudite compounds, we investigate multipole susceptibility
of a multiorbital Anderson model dynamically coupled with
Jahn-Teller phonons by using a numerical renormalization group method.
Here we take a procedure to maximize the multipole susceptibility
matrix to determine the multipole state.
When the electron-phonon coupling term is simply ignored,
it is found that the dominant multipole state is characterized by
2u octupole for the $\Gamma_{67}^-$ quartet ground state,
while the low-temperature phase is governed by magnetic fluctuations
for the $\Gamma_5^-$ doublet ground state.
When we include the coupling between $f$ electrons in degenerate
$\Gamma_{67}^{-}$ ($e_{\rm u}$) orbitals and Jahn-Teller phonons
with $E_{\rm g}$ symmetry,
the mixed multipole state with 4u magnetic and 5u octupole moments
is found to be dominant with significant 3g quadrupole fluctuations
at low temperatures.
}

\kword{Multipole, Dynamical Jahn-Teller phonons,
Sm-based filled skutterudites}

\begin{document}
\maketitle

%
%
\section{Introduction}

Recently, $f$-electron compounds including plural number of $f$ electrons
per ion have attracted renewed attention due to emergence of
exotic magnetism and unconventional superconductivity
in the research field of condensed matter physics.\cite{ASR}
In particular, filled skutterudite compounds, expressed as RT$_4$X$_{12}$
with rare-earth atom R, transition metal atom T, and pnictogen X,
have been focused, since this material group provides us
a platform for systematic research of magnetism and superconductivity
of $f^n$-electron systems with $n$$\ge$2,\cite{Sato1,Aoki1}
where $n$ denotes the number of $f$ electrons.

Among various kinds of RT$_4$X$_{12}$, interesting results have been
reported for Sm-based filled skutterudites.
SmRu$_4$P$_{12}$ is known to exhibit a metal-insulator (MI) transition
at $T_{\rm MI}$=16.5 K,\cite{Sekine1} which is confirmed to be
second order from the specific heat measurement.\cite{Matsuhira}
Further works have revealed that the MI transition occurs
in two steps.\cite{Matsuhira,Sekine2,Sekine3}
Namely, we find three states in the phase diagram for SmRu$_4$P$_{12}$
in the plane of temperature and magnetic field.\cite{Matsuhira}
The phase I ($T$$>$$T_{\rm MI}$) is paramagnetic, while the phase III
is considered to be antiferromagnetic ordered state
from NMR measurements.\cite{Fujiwara,Masaki}

Concerning the phase II between the phases I and III,
an intriguing possibility of octupole ordering has been proposed
by Yoshizawa {\it et al.} from the elastic constant measurement,
showing the breaking of time reversal symmetry
in the phase II.\cite{Yoshizawa}
Recent muon spin relaxation measurement is consistent with
this octupole scenario, since the static internal field has been
found to grow at $T_{\rm MI}$.\cite{Hachitani}
NMR measurement of $^{31}$P has also suggested that
the static internal field occurs below $T_{\rm MI}$
and the anomalous variation of the internal field is considered to
be the evidence of the ordering of octupole.\cite{Masaki,Hachitani2}

In order to understand why such exotic octupole ordering appears,
it is important to clarify the CEF ground state.
From the specific heat measurement,
for SmRu$_4$P$_{12}$ and SmOs$_4$P$_{12}$,
the CEF ground state is $\Gamma_{67}^-$ quartet,\cite{Matsuhira2}
consistent with theoretical calculation of the CEF energy level
of $T_{\rm h}$ group.\cite{Hotta1}
Note that $\Gamma_{67}^-$ in $T_{\rm h}$ group
is equal to $\Gamma_8^-$ in $O_{\rm h}$.\cite{Takegahara}
However, the CEF energy scheme seems to be changed
even among the same Sm-based filled skutterudites.
In fact, for SmFe$_4$P$_{12}$, $\Gamma_5^-$ doublet ground state
has been suggested.\cite{Matsuhira2,Nakanishi-Sm}
For SmOs$_4$Sb$_{12}$, it has been found to be $\Gamma_{67}^-$
quartet in Ref.~\citen{Sanada}, while the CEF ground state was considered
to be $\Gamma_5^-$ doublet in Ref.~\citen{Yuhasz}.
Quite recently, magnetization measurement has been performed and
the observed anisotropy seems to confirm the $\Gamma_{67}^-$ quartet
ground state in SmOs$_4$Sb$_{12}$.\cite{Aoki-Sm}

Another interesting topic on Sm-based filled skutterudites
is magnetically robust heavy-fermion behavior in
SmOs$_4$Sb$_{12}$.\cite{Sanada}
It has been suggested that this peculiar phenomenon is due to
the non-magnetic Kondo effect originating from
phononic degrees of freedom.\cite{Miyake}
Since rare-earth ion is surrounded by the cage composed of
twelve pnictogens in filled skutterudites,
there is a possibility that rare-earth ion moves around
potential minima in off-center positions inside the pnictogen cage.
This is called the rattling, which is considered to be
one of important ingredients with significant influence
on electronic properties of filled skutterudites.
As a natural extension of two-level Kondo problem,\cite{Kondo1,Kondo2}
Hattori {\it et al.} have analyzed four- and six-level Kondo models.
\cite{Hattori1,Hattori2}
Their results seem to explain the magnetically robust heavy-fermion
behavior in SmOs$_4$Sb$_{12}$.
Concerning the positions of potential minima,
neutron scattering experiment for PrOs$_4$Sb$_{12}$ has indicated
that the charge distribution of Pr ion extends in the [1,1,1] direction
at high temperatures,\cite{Kaneko}
suggesting that the eight-level Kondo model should be discussed.

Regarding the problem of rattling, there is another important
aspect of geometrical degree of freedom.
From the measurement of elastic constant,
it has been suggested that the off-center motion has
degenerate $E_{\rm g}$ symmetry.\cite{Goto}
Inspired by this suggestion, the present author has considered
a multiorbital Anderson model with a linear coupling between
degenerate $f$-electron orbitals with $e_{\rm u}$ symmetry
and dynamical Jahn-Teller phonons with $E_{\rm g}$ symmetry.\cite{Hotta2}
Numerical analysis of this model for $n$=2
has revealed that quasi-Kondo behavior occurs
due to the release of an entropy $\log 2$ of the vibronic ground state,
originating from the clockwise and anti-clockwise rotational modes
of dynamical Jahn-Teller phonons.\cite{Hotta3}
Such geometrical degree of freedom does not depend on the detail
of potential minima, since only the rotational direction
becomes important.
It is interesting to pursue a possibility of
Kondo-like behavior due to the dynamical Jahn-Teller effect
in filled skutterudites.

In this paper, we discuss multipole properties of Sm-based filled
skutterudites on the basis of the multiorbital Anderson model.
First we introduce the effective local $f$-electron model
obtained by the expansion in terms of $1/\lambda$
based on a $j$-$j$ coupling scheme,\cite{Hotta4}
where $\lambda$ is the spin-orbit coupling.
It is found that the ground state is changed between
$\Gamma_5^-$ doublet and $\Gamma_{67}^-$ quartet,
depending on Coulomb interaction and/or spin-orbit coupling.
This fact is consistent with a possible change
of the CEF ground state in SmT$_4$X$_{12}$.
Then, we numerically analyze the multiorbital Anderson model.
When the ground state is $\Gamma_{67}^-$ quartet,
the dominant multipole state is characterized by 2u octupole.
For the $\Gamma_5^-$ doublet ground state,
the low-temperature phase is governed by magnetic fluctuations.
When we consider the coupling with Jahn-Teller phonons,
the mixed multipole state with 4u magnetic and 5u octupole moments
becomes dominant at low temperatures
and quadrupole fluctuations are also significant.

The organization of this paper is as follows.
In \S 2, we introduce the effective model which describes well
the low-energy local $f$-electron states.
On the basis of the model,
we discuss the CEF energy levels of Sm$^{3+}$ from the viewpoint
of the competition between Coulomb interaction
and spin-orbit coupling.
In \S 3, the multiorbital Anderson model is defined and
we explain the quantities to be measured and
the method of calculation.
In \S 4, we show numerical results for the cases
without and with Jahn-Teller phonons.
In particular, the dominant multipole states are explained in detail.
Finally, in \S 5, we discuss relevance of the present results
to actual Sm-based filled skutterudites.
Throughout this paper, we use such units as $\hbar$=$k_{\rm B}$=1
and the energy unit is set as eV.

%
%
\section{Local $f$-electron State}

\subsection{Local effective model}

The local $f$-electron Hamiltonian is given by \cite{Hotta1}
\begin{eqnarray}
  H_{\rm loc} = H_{\rm so} + H_{\rm int} + H_{\rm CEF},
\end{eqnarray}
where $H_{\rm so}$ is the spin-orbit coupling term, given by
\begin{eqnarray}
  H_{\rm so} = \lambda \sum_{m,\sigma,m',\sigma'}
  \zeta_{m,\sigma,m',\sigma'} f_{m\sigma}^{\dag}f_{m'\sigma'}.
\end{eqnarray}
Here $\lambda$ is the spin-orbit coupling
and $f_{m\sigma}$ denotes the annihilation operator for
$f$ electron with spin $\sigma$ and angular momentum
$m$(=$-3$,$\cdots$,3) and $\sigma$=$+$1 ($-$1) for up (down) spin.
The matrix elements are expressed by
\begin{eqnarray}
  \begin{array}{l}
    \zeta_{m,\sigma,m,\sigma}=m\sigma/2,\\
    \zeta_{m+\sigma,-\sigma,m,\sigma}=\sqrt{12-m(m+\sigma)}/2,\\
  \end{array}
\end{eqnarray}
and zero for other cases.

The second term $H_{\rm int}$ indicates the Coulomb interactions
among $f$ electrons, expressed by
\begin{equation}
  H_{\rm int} = \sum_{m_1 \sim m_4 \atop \sigma_1, \sigma_2}
  I_{m_1,m_2,m_3,m_4}f_{m_1\sigma_1}^{\dag}f_{m_2\sigma_2}^{\dag}
  f_{m_3\sigma_2}f_{m_4\sigma_1},
\end{equation}
where the Coulomb integral $I_{m_1,m_2,m_3,m_4}$ is given by
\begin{eqnarray}
  I_{m_1,m_2,m_3,m_4}=\sum_{k=0}^{6} F^k c_k(m_1,m_4)c_k(m_2,m_3).
\end{eqnarray}
Note that the sum is limited by the Wigner-Eckart theorem to even values
($k$=0, 2, 4, and 6).
Here $F^k$ is the radial integral for the $k$-th partial wave,
called Slater integral or Slater-Condon parameter \cite{Slater1,Condon}
and $c_k$ is the Gaunt coefficient.\cite{Gaunt,Racah2}
The Slater-Condon parameters and the spin-orbit interaction
are determined so as to reproduce the spectra of rare-earth and actinide
ions for each value of $n$, but the Slater-Condon parameters
are considered to be distributed among 1 eV and 10 eV,
while $\lambda$ is in the order of 0.1 eV.
In this paper, for convenience, we parameterize $F^k$ as
\begin{eqnarray}
  \label{SCparam}
  F^0=10U,~F^2=5U,~F^4=3U, F^6=U,
\end{eqnarray}
where $U$ denotes the order
of the Hund's rule interaction among $f$ orbitals,
which is in the order of a few eV.

The CEF term $H_{\rm CEF}$ is given by
\begin{eqnarray}
  \label{Eq:CEF}
  H_{\rm CEF} = \sum_{m,m',\sigma} B_{m,m'}
  f_{m\sigma}^{\dag}f_{m'\sigma},
\end{eqnarray}
where $B_{m,m'}$ is determined from the table of Hutchings
for angular momentum $J$=$\ell$=3,\cite{Hutchings}
since we are now considering the potential for $f$ electron.
For filled skutterudites with $T_{\rm h}$ symmetry,\cite{Takegahara}
$B_{m,m'}$ is expressed by using three CEF parameters
$B_4^0$, $B_6^0$, and $B_6^2$ as
\begin{eqnarray}
  \begin{array}{l}
    B_{3,3}=B_{-3,-3}=180B_4^0+180B_6^0, \\
    B_{2,2}=B_{-2,-2}=-420B_4^0-1080B_6^0, \\
    B_{1,1}=B_{-1,-1}=60B_4^0+2700B_6^0, \\
    B_{0,0}=360B_4^0-3600B_6^0, \\
    B_{3,-1}=B_{-3,1}=60\sqrt{15}(B_4^0-21B_6^0),\\
    B_{2,-2}=300B_4^0+7560B_6^0,\\
    B_{3,1}=B_{-3,-1}=24\sqrt{15}B_6^2,\\
    B_{2,0}=B_{-2,0}=-48\sqrt{15}B_6^2,\\
    B_{1,-1}=-B_{3,-3}=360B_6^2.
  \end{array}
\end{eqnarray}
Note the relation of $B_{m,m'}$=$B_{m',m}$.
Following the traditional notation, we define
\begin{eqnarray}
  \begin{array}{l}
    B_4^0=Wx/F(4),\\
    B_6^0=W(1-|x|)/F(6),\\
    B_6^2=Wy/F^t(6),
  \end{array}
\end{eqnarray}
where $x$ and $y$ specify the CEF scheme for $T_{\rm h}$ point group,
while $W$ determines an energy scale for the CEF potential.
Concerning $F(4)$, $F(6)$, and $F^t(6)$, we choose
$F(4)$=15, $F(6)$=180, and $F^t(6)$=24 for $\ell$=3.\cite{Hutchings,LLW}
In actual $f$-electron materials, the magnitude of the CEF potential
is considered to be $10^{-4}$$\sim$$10^{-3}$eV.
For the calculation of the CEF energy level in this paper,
we set $W$=$-$6$\times$$10^{-4}$eV.

Although $H_{\rm loc}$ provides us correct CEF ground states
irrespective of the number of $f$ electrons and other parameters,
it is difficult to consider multipole properties and the effect of
Jahn-Teller phonons.
In order to make further steps,
it is convenient to employ the $j$-$j$ coupling scheme.\cite{Hotta5}
Namely, we accommodate $f$ electrons in the $j$=5/2 sextet
by considering the CEF potential and Coulomb interactions.
An advantage of the $j$-$j$ coupling scheme is that
we can take into account many-body effects
using standard quantum-field theoretical techniques,
since individual $f$-electron states are clearly defined.
Then, we have developed microscopic theories
for magnetism and superconductivity on the basis of the
$j$-$j$ coupling scheme by focusing on the potential roles
of $f$-electron orbitals.
\cite{Hotta0,Takimoto2,Takimoto3,Takimoto4,Hotta6,Hotta7,Onishi,Hotta8,Kubo5}
In particular, it is also possible to develop microscopic
theory for multipole ordering and fluctuations.
\cite{Kubo1,Kubo2,Kubo3,Kubo4,Kubo6,Onishi2}

However, since the sixth order CEF terms cannot be included
in the $j$=5/2 sextet due to the symmetry reason,
contributions of $B_6^0$ and $B_6^2$ are dropped
in the simple $j$-$j$ coupling scheme.
It has been a disadvantage of the $j$-$j$ coupling scheme
that the effect of $B_6^2$, which is characteristic of $T_{\rm h}$
symmetry, cannot be included.
In order to improve the situation, recently,
the modified $j$-$j$ coupling has been developed
so as to include the effect of $B_6^0$ and $B_6^2$ terms
in the order of $1/\lambda$ due to
the degenerate perturbation theory.\cite{Hotta4}
The modified $j$-$j$ coupling scheme can correctly reproduce
the CEF energy levels for $n$$\ge$2 in the realistic
intermediate coupling region with $\lambda/U$ in the order of 0.1.

In order to obtain the effective model $H_{\rm eff}$,
we treat $H_{\rm CEF}+H_{\rm int}$ as a perturbation to $H_{\rm so}$.
The details are explained in Ref.~\citen{Hotta4}.
Here we show $H_{\rm eff}$ as
\begin{eqnarray}
  \label{Heff}
  H_{\rm eff} \!=\! \sum_{\mu,\nu} {\tilde B}_{\mu,\nu}
  f_{\mu}^{\dag}f_{\nu} \!+\! \sum_{\mu_1 \sim \mu_4} \!
  {\tilde I}_{\mu_1\mu_2, \mu_3\mu_4}
  f_{\mu_1}^{\dag}f_{\mu_2}^{\dag}f_{\mu_3}f_{\mu_4},
\end{eqnarray}
where $f_{\mu}$ is the annihilation operator for
$f$ electron with angular momentum $\mu$(=$-$5/2,$\cdots$,5/2)
in the $j$=5/2 sextet.
The modified CEF potential is expressed as
\begin{equation}
  {\tilde B}_{\mu,\nu}
  ={\tilde B}^{(0)}_{\mu,\nu}+{\tilde B}^{(1)}_{\mu,\nu},
\end{equation}
where ${\tilde B}^{(0)}_{\mu,\nu}$ denotes the CEF potential for $J$=5/2
and ${\tilde B}^{(1)}_{\mu,\nu}$ is the correction in the order of
$W^2/\lambda$.
The effective interaction in eq.~(\ref{Heff}) is given by
\begin{equation}
  {\tilde I}_{\mu_1, \mu_2, \mu_3, \mu_4}=
  {\tilde I}^{(0)}_{\mu_1, \mu_2, \mu_3, \mu_4}+
  {\tilde I}^{(1)}_{\mu_1, \mu_2, \mu_3, \mu_4},
\end{equation}
where ${\tilde I}^{(0)}_{\mu_1, \mu_2, \mu_3, \mu_4}$ is
expressed by three Racah parameters, $E_0$, $E_1$, and $E_2$,
which are related to the Slater-Condon parameters as
\begin{equation}
  \label{Racah}
  \begin{array}{l}
    E_0 = F^0-(80/1225)F^2-(12/441)F^4, \\
    E_1 = (120/1225)F^2+(18/441)F^4, \\
    E_2 = (12/1225)F^2-(1/441)F^4.
  \end{array}
\end{equation}
Explicit expressions of ${\tilde I}^{(0)}$ are shown in
Ref.~\citen{Hotta5}.
On the other hand, ${\tilde I}^{(1)}_{\mu_1, \mu_2, \mu_3, \mu_4}$
is the correction term in the order of $1/\lambda$.

Here, four comments are in order.
(i) Effects of $B_6^0$ and $B_6^2$ are included as two-body
potentials in ${\tilde I}^{(1)}$.
(ii) The lowest-order energy of ${\tilde I}^{(1)}$ is
$|W|U/\lambda$.
(iii) The effective model is valid for $f$-electron compounds
with the hybridization $V$ smaller than $\lambda$.
(iv) The parameter space in which $H_{\rm eff}$ works is
determined by the conditions for the weak CEF, i.e.,
$|W|/U$$\ll$1 and $|W|U/\lambda$$\ll$$E_2$.
Concerning the value of $\lambda$,
we obtain $|W|/\lambda$$\ll$0.02
from eqs.~(\ref{SCparam}) and (\ref{Racah}).
Note that $E_2$ plays a role of the effective Hund's rule interaction
in the $j$-$j$ coupling scheme,
estimated as $E_2$$\sim$$J_{\rm H}/49$.\cite{Hotta4}
Thus, it is allowed to use $H_{\rm eff}$ even for $\lambda$
in the order of 0.1 eV, when $|W|$ is set as
a realistic value in the order of $10^{-4}$ eV
for actual $f$-electron materials.

\subsection{CEF energy schemes}

\begin{figure}[t]
\begin{center}
\includegraphics[width=8.5truecm]{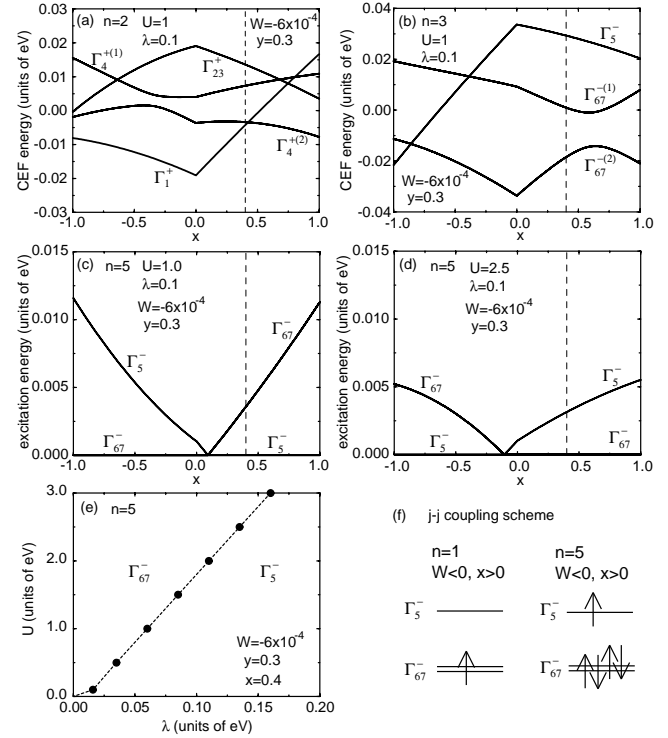}
\caption{CEF energy levels of $H_{\rm eff}$ for (a) $n$=2 and $U$=1,
(b) $n$=3 and $U$=1, (c) $n$=5 and $U$=1, and
(d) $n$=5 and $U$=2.5.
We set $\lambda$=0.1,$W$=$-$6$\times$$10^{-4}$, and $y$=0.3 for (a)-(d).
(e) Phase diagram of the CEF ground state of $H_{\rm eff}$ for $n$=5,
$W$=$-$6$\times$$10^{-4}$, $x$=0.4, and $y$=0.3.
(f) Electron configurations for $n$=1 and 5 in the $j$-$j$ coupling scheme
for $B_4^0$$<$0.}
\end{center}
\end{figure}

In order to determine the CEF parameters of filled skutterudite compounds,
first let us consider the CEF energy level for the case of $n$=2.
A typical material is PrOs$_4$Sb$_{12}$.
From experimental results on specific heat,\cite{Aoki4}
magnetization,\cite{Tayama}
and neutron scattering,\cite{Kohgi,Kuwahara,Goremychkin}
it has been confirmed that the ground state of
PrOs$_4$Sb$_{12}$ is $\Gamma_1^+$ singlet,
while the first excited state is $\Gamma_4^{+(2)}$ triplet
with very small excitation energy ($\sim$10K).

In Fig.~1(a), we show the CEF energy levels of $H_{\rm eff}$
for the case of $n$=2.
Here we set $\lambda$=0.1, $U$=1, $W$=$-$6$\times$$10^{-4}$,
and $y$=0.3, which are considered to be realistic values for
Pr-based filled skutterudites.
Here the minus sign is added in $W$ so as to be consistent
with the appearance of $\Gamma_1^+$ singlet ground state.
It should be noted that two triplet states in $O_{\rm h}$,
$\Gamma_{4}^+$ and $\Gamma_5^+$, are mixed as
$\Gamma_{4}^{+(1)}$ and $\Gamma_4^{+(2)}$ in $T_{\rm h}$.
As understood from Fig.~1(a), $x$=0.4 seems to correspond
to the actual situation of PrOs$_4$Sb$_{12}$,
since the CEF excitation energy is 1 meV,
which is consistent with experimental results.

Now we substitute the rare-earth ion from Pr$^{3+}$ ($n$=2)
to Nd$^{3+}$ ($n$=3) and Sm$^{3+}$ ($n$=5).
Here we note that in principle,
it is not necessary to change the CEF parameters
even when rare-earth ion is substituted, since the CEF potential
is given by the sum of electrostatic potentials from ligand ions.
Then, it is reasonable to use the same CEF parameters
also for NdOs$_4$Sb$_{12}$ and SmOs$_4$Sb$_{12}$.
In actuality, the electrostatic potential may be changed
depending on the ion radius of R and/or the kinds of T and X.
In this sense, $x$, $y$, and the absolute value of $|W|$ may be
changed among filled skutterudite compounds,
but there is no reason to change the sign of $W$.
Throughout this paper, we keep $W$ as negative.

Let us discuss the CEF states for $n$=3.\cite{HottaICM}
In Fig.~1(b), we show the CEF energy levels of $H_{\rm loc}$
for $n$=3 by using the same parameters as in Fig.~1(a).
At $x$=0.4, the ground state is $\Gamma_{67}^-$ quartet
and the excited state is another $\Gamma_{67}^-$ quartet,
consistent with the experimental result for NdOs$_4$Sb$_{12}$.\cite{Ho}
As for the CEF excitation energy,
it is 0.02eV in the present calculation,
while it has been experimentally estimated as 220K.\cite{Ho}
We can obtain the CEF energy for NdOs$_4$Sb$_{12}$
consistent with the experimental result
by using the CEF parameters of PrOs$_4$Sb$_{12}$.

In Fig.~1(c), we show the CEF states for $n$=5
by using the same parameters as in Fig.~1(a).
Since we obtain two multiplets, $\Gamma_5^-$ and $\Gamma_{67}^-$,
for $n$=5, it is convenient to draw the excitation energy.
At $x$=0.4, the ground state is $\Gamma_{5}^-$ doublet,
while the experimental result for SmOs$_4$Sb$_{12}$ has
suggested $\Gamma_{67}^-$ quartet ground state.\cite{Sanada}
Here we note that the values of $U$ and $\lambda$ can
be different for different rare-earth ions.
In the cases of $n$=2 and $n$=3, even if $U$ and
$\lambda$ are changed within the realistic range,
the CEF energy scheme does not change qualitatively.
However, this is not the case for $n$=5.

In Fig.~1(d), we show the results of
the CEF energy levels for $U$=2.5 and $\lambda$=0.1
by using the same CEF parameters as in Fig.~1(c).
It is observed that 
the $x$ dependence is reversed from that in Fig.~1(c),
suggesting that the CEF ground state is changed,
even though we do not change the CEF parameters.
At $x$=0.4, the ground state is $\Gamma_{67}^-$ quartet
and the excitation energy is about 3 meV in the present
calculation, which is comparable with the experimental
value of 19K.\cite{Sanada}

We note that the conversion of the ground state by the
change of $U$ (or $\lambda$) is $not$ due to the approximation
of $H_{\rm eff}$, but an intrinsic feature of $f$ electrons,
since we could find the same results when we
diagonalize simultaneously CEF potential,
Coulomb interaction, and spin-orbit coupling.\cite{Hotta4}
By changing $U$ and $\lambda$, we depict the phase diagram
for the CEF ground state for the case of $n$=5 and $x$=0.4,
as shown in Fig.~1(e).
Note that in the region of very small $\lambda$,
$H_{\rm eff}$ does not work, but it is still available
for $\lambda$$\gg$0.015 for the present parameter choice.
Roughly speaking, for large $\lambda$ and small $U$,
the ground state is $\Gamma_{67}^-$ quartet,
while for small $\lambda$ and large $U$,
$\Gamma_{5}^-$ doublet is the ground state.

In order to understand the change of the CEF ground state,
it is useful to consider the two limiting situations,
$\lambda$$\ll$$U$ and $\lambda$$\gg$$U$,
corresponding to the $LS$ and $j$-$j$ coupling schemes, respectively.
First we consider the case of the $LS$ coupling scheme.
The relevant CEF parameter is $B_4^0$, which is expressed as
$B_4^0$=$A_4$$\beta_J(n) \langle r^4 \rangle$,
where $A_4$ is a parameter depending on materials,
$\beta_J(n)$ is the so-called Stevens factor,
appearing in the method of operator equivalent,\cite{Stevens}
and $\langle r^4 \rangle$ indicates the radial average
concerning the local $f$-electron wavefunction.
When we assume that $A_4$ is invariant in the same material group
and $\langle r^4 \rangle$ does not depend on the $f$-electron number,
the CEF parameter is basically determined by $\beta_J(n)$,
which depend on $n$, $U$, and $\lambda$.\cite{Hotta4}

In the $LS$ coupling scheme,
$\beta_{5/2}(1)$=(11/7)$\beta_{\ell}$
and $\beta_{5/2}(5)$=(13/21)$\beta_{\ell}$,
where $\beta_{\ell}$=$\beta_{3}(1)$=2/495 for $\ell$=3.
Note that $\beta_{5/2}(n)$ is different between the cases of
$n$=1 and 5 even for the same $J$=5/2,
since $L$=3 and $S$=1/2 for $n$=1, while $L$=5 and $S$=5/2 for $n$=5,
where $L$ and $S$ are total angular momentum and spin momentum,
respectively.
Namely, in the $LS$ coupling scheme,
$B_4^0$ should have the same sign for $n$=1 and 5.
The CEF ground state for $n$=1 is $\Gamma_{5}^-$ doublet
for $B_4^0$$>$0, while it is $\Gamma_{67}^-$ quartet for $B_4^0$$<$0.
Thus, in the $LS$ coupling scheme, it is easy to understand that
the ground state for $n$=5 is the $\Gamma_{67}^-$ quartet
in the region of $x$$>$0 for the present parameterization of
$B_4^0$=$Wx/15$ with negative $W$.

Next let us consider another limiting case corresponding to
the $j$-$j$ coupling scheme, in which we simply accommodate
electrons in the one-electron potentials,\cite{Hotta5}
as shown in Fig.~1(f).
For $x$$>0$ with $W$$<$0, i.e., negative $B_4^0$,
we put five electrons in the level scheme of lower $\Gamma_{67}^-$
and higher $\Gamma_{5}^-$.
Thus, we immediately understand that the ground state becomes
$\Gamma_{5}^-$ doublet for $n$=5 in the $j$-$j$ coupling scheme,
when $\Gamma_{67}^-$ quartet is the ground state for $n$=1.
In fact, in the $j$-$j$ coupling scheme,
we obtain $\beta_{5/2}(1)$=(11/7)$\beta_{\ell}$
and $\beta_{5/2}(5)$=($-$11/7)$\beta_{\ell}$.
The sign of $B_4^0$ is different between the cases
of $n$=1 and 5 in the $j$-$j$ coupling scheme.
When we change the values of $U$ and/or $\lambda$ from
the $LS$ to $j$-$j$ coupling regions,
we can find the smooth change of $\beta_{5/2}(5)$ from
(13/21)$\beta_{\ell}$ to ($-$11/7)$\beta_{\ell}$,
as found in Ref.~\citen{Hotta4}

The present result suggests a ``non-CEF'' scenario
to understand the change of the CEF ground state in
Sm-based filled skutterudites.
As mentioned in the introduction,
for SmRu$_4$P$_{12}$ and SmOs$_4$P$_{12}$,
the CEF ground state is $\Gamma_{67}^-$ quartet,
while the ground state of SmFe$_4$P$_{12}$ is
$\Gamma_5^-$ doublet.
In order to explain the change of the ground state
within the $LS$ coupling scheme,
it is necessary to change the sign of $B_4^0$.
The magnitude of the CEF potential may be changed
due to the difference in the ion radius of R and
the substitution of T and/or X in RT$_4$X$_{12}$,
but it seems difficult to imagine that the sign of
the CEF potential is changed among SmT$_4$X$_{12}$.
As understood from Fig.~1(e),
it is observed that the ground state conversion occurs
around the region for realistic values of $\lambda$ and $U$.
Namely, due to the change in $U$ and/or $\lambda$,
the ground state is easily interchanged for
actual Sm materials.

%
%
\section{Model and Method}

\subsection{Multiorbital Anderson Hamiltonian}

From the band-structure calculations,
it has been revealed that the main conduction band of
filled skutterudites is $a_{\rm u}$ with xyz symmetry,\cite{Harima1}
which is hybridized with $f$ electrons in the $\Gamma_5^-$ state
with $a_{\rm u}$ symmetry.
In order to specify the $f$-electron state,
we introduce ``orbital'' index which distinguishes
three kinds of the Kramers doublets,
two $\Gamma_{67}^-$ and one $\Gamma_5^-$.
Here ``a'' and ``b'' denote the two $\Gamma_{67}^-$'s and ``c''
indicates the $\Gamma_5^-$.

Then, the multiorbital Anderson model is given by
\begin{equation}
  \label{Ham}
  \begin{split}
    H & = \sum_{\mib{k}\sigma}
    \varepsilon_{\mib{k}} c_{\mib{k}\sigma}^{\dag} c_{\mib{k}\sigma}
     +  \sum_{\mib{k}\sigma}
    (V c_{\mib{k}\sigma}^{\dag}f_{{\rm c}\sigma}+{\rm h.c.}) \\
     & + H_{\rm eff}  +  H_{\rm eph},
  \end{split}
\end{equation}
where $\varepsilon_{\mib{k}}$ is the dispersion of $a_{\rm u}$
conduction electrons with $\Gamma_5^-$ symmetry,
$f_{\gamma\sigma}$ is the annihilation operator
of $f$ electrons on the impurity site with pseudospin $\sigma$
and orbital $\gamma$,
$c_{\mib{k}\sigma}$ is the annihilation
operator for conduction electrons with momentum $\mib{k}$ and
pseudo-spin $\sigma$,
and $V$ is the hybridization between conduction and $f$ electrons
with $a_{\rm u}$ symmetry.
Throughout this paper, we set $V$=0.05 eV.
Note that the energy unit of $H$ is half of the bandwidth of
the conduction band,
which is considered to be in the order of 1 eV,
since the bandwidth has been typically estimated as 2.7 eV
for PrRu$_4$P$_{12}$ \cite{Harima2}.
Thus, the energy unit of $H$ is taken as eV.
To set the local $f$-electron number as $n$=5,
we adjust the $f$-electron chemical potential.

The last term in eq.~(\ref{Ham}) denotes 
the electron-phonon coupling.
Here, the effect of $E_{\rm g}$ rattling is included as
relative vibration of surrounding atoms.
We remark that localized $\Gamma_{67}^-$ orbitals with
$e_{\rm u}$ symmetry have linear coupling with JT phonons with
$E_{\rm g}$ symmetry, since the symmetric representation of
$e_{\rm u}$$\times$$e_{\rm u}$ includes $E_{\rm g}$.
Then, $H_{\rm eph}$ is given by
\begin{equation}
  \label{Heph}
  \begin{split}
   H_{\rm eph} & =g (Q_2 \tau_x + Q_3 \tau_z)+(P_2^2+P_3^2)/2 \\
   & +(\omega^2/2)(Q_2^2+Q_3^2) + b(Q_3^3-2Q_2^2Q_3),
  \end{split}
\end{equation}
where $g$ is the electron-phonon coupling constant,
$Q_2$ and $Q_3$ are normal coordinates for $(x^2-y^2)$-
and $(3z^2-r^2)$-type JT phonons, respectively,
$P_2$ and $P_3$ are corresponding canonical momenta,
$\tau_{x}$=
$\sum_{\sigma}(f_{{\rm a}\sigma}^{\dag}f_{{\rm b}\sigma}
+f_{{\rm b}\sigma}^{\dag}f_{{\rm a}\sigma})$,
$\tau_{z}$=
$\sum_{\sigma}(f_{{\rm a}\sigma}^{\dag}f_{{\rm a}\sigma}
-f_{{\rm b}\sigma}^{\dag}f_{{\rm b}\sigma})$,
$\omega$ is the frequency of local JT phonons,
and $b$ indicates the cubic anharmonicity.
Note that the reduced mass for JT modes is set as unity.
Here we introduce non-dimensional
electron-phonon coupling constant $\alpha$ and
the anharmonic energy $\beta$ as $\alpha$=$g^2/(2\omega^3)$
and $\beta$=$b/(2\omega)^{3/2}$, respectively.

\subsection{Multipole Susceptibility}

In order to clarify the magnetic properties at low temperatures,
we usually discuss the magnetic susceptibility,
but in more general,
it is necessary to consider the susceptibility of multipole moments
such as dipole, quadrupole, and octupole.\cite{Hotta9}
The multipole operator is given in the second-quantized form as
\begin{equation}
   \label{eqX}
   X_{\gamma} = \sum_{\mu,\nu}
   (X_{\gamma})_{\mu\nu} f_{\mu}^{\dag}f_{\nu},
\end{equation}
where $X$ denotes the symbol of multipole with the symmetry of
$\Gamma_{\gamma}$ and $\gamma$ indicates a set of indices
for the irreducible representation.
For $j$=5/2, we can define multipole operators
up to rank 5 in general, but we are interested in
multipole properties from the $\Gamma_8$ quartet.
Thus, we concentrate on multipole moments up to rank 3.

Now let us show explicit forms for multipole operators
following Ref.~\citen{Shiina}. See also Ref.~\citen{Hotta9}.
As for dipole moments with $\Gamma_{\rm 4u}$ symmetry,
we express the operators as
\begin{equation}
 J_{{\rm 4u}x}=J_x,~J_{{\rm 4u}y}=J_y,~J_{{\rm 4u}z}=J_z,
\end{equation}
where $J_x$, $J_y$, and $J_{z}$ are three angular momentum operators
for $j$=5/2, respectively.
Concerning quadrupole moments, they are classified into
$\Gamma_{\rm 3g}$ and $\Gamma_{\rm 5g}$.
We express $\Gamma_{\rm 3g}$ quadrupole operators as
\begin{equation}
\begin{array}{l}
 O_{{\rm 3g}u} = (2J_z^2-J_x^2-J_y^2)/2, \\
 O_{{\rm 3g}v} = \sqrt{3}(J_x^2-J_y^2)/2.
\end{array}
\end{equation}
For the $\Gamma_{\rm 5g}$ quadrupole, we have the three operators 
\begin{equation}
\begin{array}{l}
 O_{{\rm 5g}\xi} = \sqrt{3} \, \overline{J_yJ_z}/2,\\
 O_{{\rm 5g}\eta} = \sqrt{3} \, \overline{J_zJ_x}/2,\\
 O_{{\rm 5g}\zeta} = \sqrt{3} \, \overline{J_xJ_y}/2,
\end{array}
\end{equation}
where the bar denotes the operation of taking all possible
permutations in terms of cartesian components.

Regarding octupole moments, there are three types as
$\Gamma_{\rm 2u}$, $\Gamma_{\rm 4u}$, and
$\Gamma_{\rm 5u}$.
We express the $\Gamma_{\rm 2u}$ octupole as
\begin{equation}
  T_{\rm 2u}=\sqrt{15} \, \overline{J_xJ_yJ_z}/6.
\end{equation}
For the $\Gamma_{\rm 4u}$ octupole, we express the operators as
\begin{equation}
\begin{array}{l}
 T_{{\rm 4u}x}=(2J_x^3-\overline{J_xJ_y^2}-\overline{J_xJ_z^2})/2,\\
 T_{{\rm 4u}y}=(2J_y^3-\overline{J_yJ_z^2}-\overline{J_yJ_x^2})/2,\\
 T_{{\rm 4u}z}=(2J_z^3-\overline{J_zJ_x^2}-\overline{J_zJ_y^2})/2,
\end{array}
\end{equation}
while the $\Gamma_{\rm 5u}$ octupole operators are given by
\begin{equation}
\begin{array}{l}
 T_{{\rm 5u}x}=\sqrt{15}(\overline{J_xJ_y^2}-\overline{J_xJ_z^2})/6,\\
 T_{{\rm 5u}y}=\sqrt{15}(\overline{J_yJ_z^2}-\overline{J_yJ_x^2})/6,\\
 T_{{\rm 5u}z}=\sqrt{15}(\overline{J_zJ_x^2}-\overline{J_zJ_y^2})/6.
\end{array}
\end{equation}
We redefine the multipole moments so as to satisfy the orthonormal condition
Tr$(X_{\gamma}X_{\gamma'})$=$\delta_{\gamma\gamma'}$ \cite{Kubo7},
where $\delta_{\gamma\gamma'}$ is the Kronecker's delta.

In principle, the multipole susceptibility can be evaluated
in the linear response theory,\cite{Hotta9}
but we should note that the multipole moments
belonging to the same symmetry can be mixed in general.
In order to determine the coefficient of such mixed multipole moment,
it is necessary to find the optimized multipole state which
maximizes the susceptibility.
Namely, we define the multipole operator as
\begin{equation}
  M = \sum_{\gamma} p_{\gamma}X_{\gamma},
\end{equation}
where $p_{\gamma}$ is determined by the eigenstate
with the maximum eigenvalue of the susceptibility matrix, given by
\begin{eqnarray}
  \label{sus}
  \chi_{\gamma\gamma'} \! = \!
  \frac{1}{Z} \sum_{n,m}
  \frac{e^{-E_n/T}-e^{-E_m/T}}{E_m-E_n} \!
  \langle n | X_{\gamma} | m \rangle
  \langle m | X_{\gamma'} | n \rangle,
\end{eqnarray}
where $E_n$ is the eigenenergy for the $n$-th eigenstate
$|n\rangle$ and $Z$ is the partition function given by
$Z$=$\sum_n e^{-E_n/T}$.

\subsection{Method}

In order to evaluate multipole susceptibilities, here we employ
a numerical renormalization group (NRG) method,\cite{NRG1,NRG2}
in which momentum space is logarithmically discretized
to include efficiently the conduction electrons near the Fermi energy
and the conduction electron states
are characterized by ``shell'' labeled by $N$.
The shell of $N$=0 denotes an impurity site described by
the local Hamiltonian.
The Hamiltonian is transformed into the recursion form as
\begin{eqnarray}
  H_{N+1} = \sqrt{\Lambda}H_N+t_N \sum_\sigma
  (c_{N\sigma}^{\dag}c_{N+1\sigma}+c_{N+1\sigma}^{\dag}c_{N\sigma}),
\end{eqnarray}
where $\Lambda$ is a parameter for logarithmic discretization,
$c_{N\sigma}$ denotes the annihilation operator of conduction electron
in the $N$-shell, and $t_N$ indicates ``hopping'' of electron between
$N$- and $(N+1)$-shells, expressed by
\begin{eqnarray}
  t_N=\frac{(1+\Lambda^{-1})(1-\Lambda^{-N-1})}
  {2\sqrt{(1-\Lambda^{-2N-1})(1-\Lambda^{-2N-3})}}.
\end{eqnarray}
The initial term $H_0$ is given by
\begin{eqnarray}
  H_0=\Lambda^{-1/2}[H_{\rm eff} + H_{\rm eph} + \sum_{\sigma}
  V(c_{0\sigma}^{\dag}f_{{\rm c}\sigma}
  +f_{{\rm c}\sigma}^{\dag}c_{0\sigma})].
\end{eqnarray}
The component of multipole susceptibility eq.~(\ref{sus})
is evaluated by using the renormalized state.
The free energy $F$ for $f$ electron is evaluated by
\begin{eqnarray}
   F = -T \lim_{N \rightarrow \infty}
   \Bigl[ \ln {\rm Tr} e^{-H_N/T}
   - \ln {\rm Tr} e^{-H_N^0/T} \Bigr].
\end{eqnarray}
Note that the temperature $T$ is defined as
$T$=$\Lambda^{-(N-1)/2}$ in the NRG calculation.
The entropy $S_{\rm imp}$ is obtained by
$S_{\rm imp}$=$-\partial F/\partial T$
and the specific heat $C_{\rm imp}$ is evaluated by
$C_{\rm imp}$=$-T\partial^2 F/\partial T^2$.
In this paper, $\Lambda$ is set as $5$ and we keep $3000$
low-energy states for each renormalization step.
The phonon basis for each JT mode
is truncated at a finite number $N_{\rm ph}$,
which is set as $N_{\rm ph}$=20.

%
%
\section{Results}

\subsection{Without Jahn-Teller phonons}

\begin{figure}[t]
\centering
\includegraphics[width=8.5truecm]{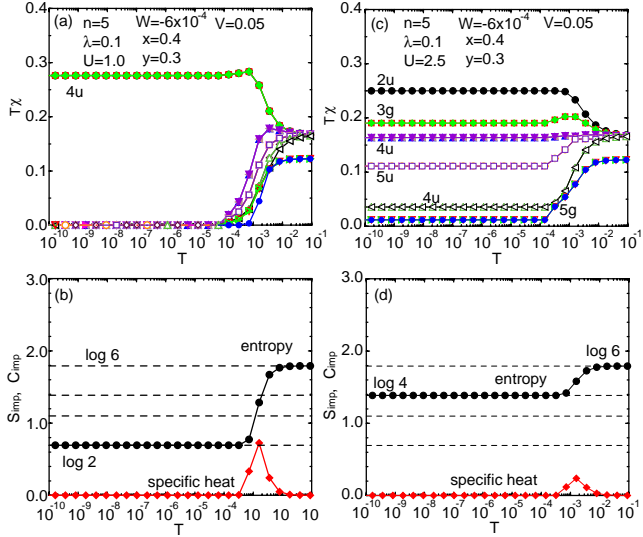}
\caption{(a) $T\chi$ and (b) $S_{\rm imp}$ and $C_{\rm imp}$
vs. temperature for $\lambda$=0.1 and $U$=1.0.
(c) $T\chi$ and (d) $S_{\rm imp}$ and $C_{\rm imp}$
vs. temperature for $\lambda$=0.1 and $U$=2.5.}
\end{figure}

First let us consider the case without $H_{\rm eph}$.
In Fig.~2(a), we show the multipole susceptibility $\chi$
for $U$=1 and $\lambda$=0.1 with the $\Gamma_{5}^-$ ground state.
At low temperatures, we find the 4u magnetic state and other
multipoles are suppressed.
Note that the 4u magnetic moment is expressed as
\begin{equation}
  \label{mom}
  M_{4ua} = p_a J_{{\rm 4u}a}+q_a T_{{\rm 4u}a},
\end{equation}
for $a$=$x$, $y$, and $z$,
since the moments belonging in the same symmetry are mixed
in general.\cite{Kubo7}
For the present parameters,
we find $p_a$=0.326 and $q_a$=$-$0.946.
As shown in Fig.~2(b), we find a residual entropy $\log 2$ of
$\Gamma_5^-$ doublet, but it is not the true ground state.
For instance, when we increase the hybridization $V$,
the 4u magnetic moment is screened
even in the present temperature range.
In any case, for the case of $\Gamma_5^-$ ground state,
there are no significant higher-order multipoles.
We find the standard Kondo physics,
when the extra coupling to JT phonons is not considered.

On the other hand, the low-temperature state is drastically
changed, when the ground state is $\Gamma_{67}^-$ quartet,
as shown in Fig.~2(c).
After the diagonalization of the susceptibility matrix,
we obtain that the dominant multipole is purely given by 2u octupole.
The second and third multipole states have 3g quadrupole
and 4u magnetic moments, respectively.
The 4u moment is expressed by eq.~(\ref{mom}) with
$p_a$=0.560 and $q_a$=0.828.
The fourth multipole state is given by 5u octupole,
but in the case of $n$=5, 4u moment is not mixed with 5u,
in contrast to the case of $n$=3.\cite{HottaICM}
As explained in the previous section for the CEF states of $n$=2,
4u and 5u states in $O_{\rm h}$ symmetry are mixed in $T_{\rm h}$.
Thus, the mixture of 4u and 5u moments is one of characteristic points
of filled skutterudites with $T_{\rm h}$ symmetry.
However, this mixing occurs due to the CEF term of $B_6^2$,
which does not appear in the $J$=5/2 space for $n$=1 and 5.
Thus, in Sm-based filled skutterudites,
the mixing of 4u and 5u moments does not occur
within an electronic model.
In Fig.~2(c), we also find another 4u state and 5g quadrupole.

In the case of the $\Gamma_{67}^-$ ground state,
since this quartet carries all the multipole moments up to rank 3,
we find the significant values for all moments.
As shown in Fig.~2(d), we find a residual entropy of
$\log 4$, corresponding to the localized $\Gamma_{67}^-$ quartet,
since we have considered the hybridization between $a_{\rm u}$
conduction band and $\Gamma_5^-$ state.
In actuality, there should exist a finite
hybridization between $e_{\rm u}$ conduction bands and
$\Gamma_{67}^-$ states, even if the value is not large compared
with that between $a_{\rm u}$ conduction and $\Gamma_5^-$ electrons.
Thus, the entropy of $\log 4$ should be eventually released
at low temperatures in the actual situation.

\subsection{With Jahn-Teller phonons}

Next we include the effect of dynamical JT phonons,
but before proceeding to the numerical results,
let us briefly review the effect of JT phonons
in the adiabatic approximation by following Ref.~\citen{Hotta3}.
Note that in actuality, the potential is not static,
but it dynamically changes to follow the electron motion.
For $\beta$=0, the potential is continuously degenerate
along the circle of the bottom of the Mexican-hat potential.
Thus, we obtain double degeneracy in the vibronic state
concerning the rotational JT modes
along clockwise and anti-clockwise directions.
When a temperature becomes lower than a characteristic energy
$T^*$, which is related to a time scale to turn
the direction of rotational JT modes,
the entropy $\log 2$ should be eventually released,
leading to Kondo-like behavior,
since the specific rotational direction disappears
due to the average over the long enough time.

In Figs.~3(a) and 3(b), we show multipole susceptibilities, entropy,
and specific heat for $\omega$=0.3, $\alpha$=0.5, and $\beta$=0.0.
Note that here we suppress the cubic anharmonicity.
We find that all the multipole moments are still active,
although the dominant multipole is changed from 2u to 4u.
In this sense, the results do not seem to be qualitatively
changed between Figs.~2(c) and 3(a),
in spite of the JT active situation.
We find a residual entropy of $\log 4$ in Fig.~3(b).
In Fig.~3(c), we show the temperature dependence of
average displacements.
For the whole temperature range,
we find finite values of
$\sqrt{\langle Q_2^2 \rangle}$ and $\sqrt{\langle Q_3^2 \rangle}$,
while $\langle Q_2 \rangle$=$\langle Q_3 \rangle$=0.
Namely, JT vibrations occur around the origin without displacements.
As the temperature is decreased, such vibrations
are expected to be suppressed, but in the temperature range,
we cannot find it.
Rather, due to the isotropic JT vibrations,
orbital fluctuations are still active in addition to
magnetic fluctuations, leading to active multipole fluctuations.

As explained in Ref.~\citen{Hotta3},
when we include the effect of anharmonicity,
three potential minima appear in the bottom of the JT potential
in the adiabatic approximation.
Since the rotational mode should be changed to the quantum tunneling
among three potential minima at low temperatures,
the frequency is effectively reduced in the factor of
$e^{-\delta E/\omega}$,
where $\delta E$ is the potential barrier.
Then, we expect to observe the quasi-Kondo behavior
even in the present temperature range, when we include the
effect of cubic anharmonicity.

\begin{figure}[t]
\centering
\includegraphics[width=8.5truecm]{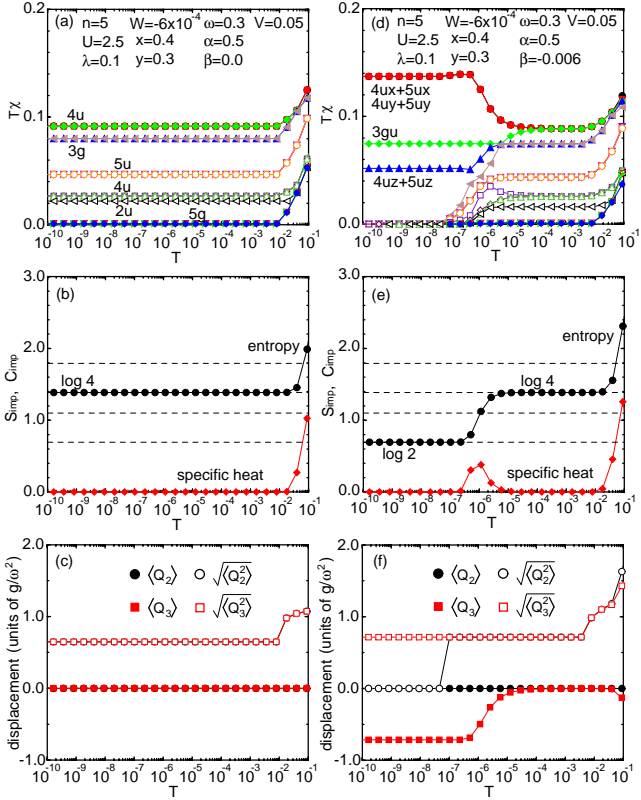}
\caption{(a) $T\chi$, (b) $S_{\rm imp}$ and $C_{\rm imp}$,
and (c) $\langle Q_i \rangle$ and $\sqrt{\langle Q_i^2 \rangle}$
($i$=2 and 3) vs. temperature for $\lambda$=0.1, $U$=2.5,
and $\beta$=0.
(d) $T\chi$, (e) $S_{\rm imp}$ and $C_{\rm imp}$,
and (f) $\langle Q_i \rangle$ and $\sqrt{\langle Q_i^2 \rangle}$
($i$=2 and 3) vs. temperature for $\lambda$=0.2, $U$=1.0,
and $\beta$=$-$0.006.
}
\end{figure}

In Figs.~3(d) and 3(e), we show multipole susceptibilities, entropy,
and specific heat for $\omega$=0.3, $\alpha$=0.5, and $\beta$=$-$0.006.
For $T$$>$$10^{-5}$, multipole susceptibilities are quite similar to
Fig.~3(a).
However, when temperature is decreased, the mixed multipole state
with 4u magnetic and 5u octupole moments becomes dominant.
This mixed moment is expressed as
\begin{equation}
  M_{a} = p_a J_{{\rm 4u}a}+q_a T_{{\rm 4u}a} + r_a T_{{\rm 5u}a},
\end{equation}
for $a$=$x$ and $y$.
In the low-temperature region,
we obtain $p_a$=0.761, $q_a$=0.428, and $r_a$=$-$0.488.
Due to the effect of JT phonons, we find a significant contribution
of 5u octupole.
It is emphasized that the 5u component first appears in the present case
for $n$=5.
The secondary moment is given by 3g$u$ quadrupole and
the third state is given by
\begin{equation}
  M_{z} = p_z J_{{\rm 4u}z}+q_z T_{{\rm 4u}z} + r_z T_{{\rm 5u}z},
\end{equation}
where $p_z$=0.67, $q_z$=$-$0.739, and $r_z$=0.0.
Note that $r_z$ is found to be zero within the numerical precision,
but it is difficult to conclude that it vanishes exactly.

In Fig.~3(e), around at $T$=$10^{-6}$, we find a peak in the specific heat,
since an entropy of $\log 2$ is released.
As mentioned above, this is considered to be quasi-Kondo behavior,
originating from the suppression of the rotational mode
of dynamical JT phonons.\cite{Hotta3}
In this case, an entropy of $\log 2$ concerning orbital degree of freedom
coupled with JT phonons is released,
while there still remains spin degree of freedom in
the localized $\Gamma_{67}^-$ quartet.
In fact, at low temperatures, magnetic susceptibility becomes dominant.

In Fig.~3(f), we show the temperature dependence of average displacements.
For $T$$>$$10^{-6}$, we find $\sqrt{\langle Q_2^2 \rangle}$$\ne$0
and $\sqrt{\langle Q_3^2 \rangle}$$\ne$0,
suggesting that both $Q_2$ and $Q_3$ modes are active.
This is consistent with the finite values of susceptibilities
for $O_{{\rm 3g}u}$ and $O_{{\rm 3g}v}$.
Note that the displacement is not considered to occur,
since $\langle Q_2 \rangle$=$\langle Q_3 \rangle$=0.
In the low-temperature region, after the release of
an entropy $\log 2$ of the rotational degree of freedom,
we find $\sqrt{\langle Q_2^2 \rangle}$=$\langle Q_2 \rangle$=0,
while $\sqrt{\langle Q_3^2 \rangle}$=$|\langle Q_3 \rangle|$$\ne$0,
indicating that only $Q_3$-type JT vibration is active
with finite displacement.
This is also consistent with the result that
$\chi_{{\rm 3g}u}$ remains at low temperatures,
since the vibration mode is fixed as $Q_3$-type,
after the quasi-Kondo phenomenon occurs.
Namely, the JT vibrations become anisotropic for $T$$<$$T^*$,
while they are isotropic for $T$$>$$T^*$,
as mentioned in Ref.~\citen{Hotta3}.

For the case with JT phonons, we have shown only the results
for the $\Gamma_{67}^-$ ground state.
When we evaluate the multipole susceptibility, specific heat,
entropy, average displacement for the $\Gamma_{5}^-$ ground state,
we find the results quite similar to Figs.~3.
In the present parameters, the ground state character is masked
due to the effect of JT phonons.
If we increase the CEF excitation energy,
we could find different results from those of Fig.~3 for the $\Gamma_5^-$
ground state, but it is not so important in the context of
Sm-based filled skutterudites with the CEF excitation in
the order of 10K.

%
%
\section{Discussion and Summary}

In this paper, we have examined the multipole state
of Sm-based filled skutterudites on the basis of
the multiorbital Anderson model with the use of
the NRG method.
As mentioned in the introduction,
a possibility of octupole ordering has been discussed
experimentally in SmRu$_4$P$_{12}$.\cite{Yoshizawa,Hachitani,Masaki}
In general, there are two possibilities as
2u and 5u octupoles.
The 2u octupole has been found to be dominant
in the electronic model,
when we do not include the coupling between
$f$ electrons and JT phonons.
On the other hand, the 5u octupole becomes significant in the mixed
state with 4u magnetic moment,
when we consider the coupling to JT phonons.

In the experiment, Yoshizawa {\it et al.} have suggested the 5u octupole
from the elastic constant measurement.\cite{Yoshizawa}
The isotropic 2u octupole seems to contradict with experimental results.
Note here that the pure electronic model cannot stabilize the 5u octupole,
since there is no mixing between 4u and 5u moments for the case of $n$=5
with the ground state multiplet characterized by $J$=5/2.
To obtain the mixed multipoles of 4u magnetic and 5u octupole,
it is essentially important to consider the electron-phonon coupling.
It is not yet clarified that the dominant phonon mode in filled skutterudites
is Jahn-Teller type with $E_{\rm g}$ symmetry,
but in any case, the effect of rattling seems to play a crucial role
for the appearance of the octupole state.
It may be interesting to carry out an experiment to detect
a coupling between rattling and $f$-electron state in SmRu$_4$P$_{12}$.

Concerning the mechanism of
magnetically robust heavy-fermion phenomena observed
in SmOs$_4$Sb$_{12}$ \cite{Sanada},
a potential role of phonons has been pointed out from the viewpoint
of the Kondo effect with non-magnetic origin \cite{Miyake}.
In this context, the quasi-Kondo behavior due to the dynamical
JT phonons may be a possible candidate to understand
magnetically robust heavy-fermion phenomenon.
Further investigations are required, in particular, concerning
an experimental method to detect the Kondo-like behavior
due to Jahn-Teller phonons.
It is one of important future issues.

In summary, we have discussed the multipole state for the case of
$n$=5 by analyzing the multipole Anderson model with the use of
the NRG technique.
When we do not consider the coupling between JT phonons and
$f$ electrons in $\Gamma_{67}^-$ quartet, we have found that
the dominant multipole moment is 2u octupole.
When the coupling with JT phonons is switched and
the cubic anharmonicity is included,
the dominant multipole state is characterized by
the mixture of 4u magnetic and 5u octupole moments,
with significant fluctuations of quadrupole.
We have also observed the quasi-Kondo behavior
due to the entropy release concerning the rotational JT mode.

\section*{Acknowledgement}

The author thanks Y. Aoki, H. Harima, K. Kubo, H. Onishi, Y. Nakanishi,
and M. Yoshizawa for discussions.
This work has been supported by a Grant-in-Aid for Scientific Research
in Priority Area ``Skutterudites'' under the contract No.~18027016
from the Ministry of Education, Culture, Sports, Science, and
Technology of Japan.
The author has been also supported by a Grant-in-Aid for
Scientific Research (C) under the contract No.~18540361
from Japan Society for the Promotion of Science.
The computation in this work has been done using the facilities
of the Supercomputer Center of Institute for Solid State Physics,
University of Tokyo.


\end{document}